\documentstyle[12pt]{article}
\newcommand{\be}{\begin{equation}}
\newcommand{\ee}{\end{equation}}
\newcommand{\bea}{\begin{eqnarray}}
\newcommand{\eea}{\end{eqnarray}}
\newcommand{\ba}{\begin{array}}
\newcommand{\ea}{\end{array}}
\newcommand{\beas}{\begin{eqnarray*}}
\newcommand{\eeas}{\end{eqnarray*}}
\newcommand{\bes}{\begin{equation*}}
\newcommand{\ees}{\end{equation*}}

\newcommand{\nn}{\nonumber}

\newcommand{\f}{\frac}

\def\cL{{\cal L}}

\def\i2           {\mbox{$\frac{i}{2}$}}

\def\al           {\alpha}

\def\bet           {\beta}

\def\ch           {\chi}

\def\del           {\delta}

\def\ep           {\epsilon}

\def\et           {\eta}

\def\ga           {\gamma}

\def\la           {\lambda}

\def\pf            {{\rm Pf }}
\def\ph           {\phi}

\def\ps           {\psi}

\def\th{\theta}

\def\pl           {\partial}

\def \tf {{\tilde F}}

\begin{document}

\title{ Topological DBI actions and nonlinear instantons }
\author{ A. Imaanpur \thanks{Email: aimaanpu@theory.ipm.ac.ir} \\
{\small {\em Department of Physics, School of Sciences,}}\\
{\small {\em Tarbiat Modares University, P.O.Box 14155-4838, 
Tehran, Iran, and}} \\ 
{\small {\em Institute for Studies in Theoretical Physics and Mathematics}} \\
{\small {\em P.O.Box 19395-5531, Tehran, Iran}}}

\maketitle

\begin{abstract}

We consider Euclidean D4 and D6-branes filling 
the whole ${\bf R}^4$ and ${\bf R}^6$ space, respectively. 
In both cases, with a constant background B-field turned on 
for D4-branes,   
we propose actions which are  
the same as the DBI actions up to some constant or total 
derivative terms. These extra terms allow us to write the action 
as a square of nonlinear instanton equations. 
As such, the actions can easily be 
supersymmetrized using the methods of topological field theory. 

\end{abstract}

\section{Introduction}
D-branes in string theory are effectively described by the Dirac-Born-Infeld 
action. The action has also a  supersymmetric extension 
which consists of two parts; the Dirac-Born-Infeld (DBI) part, and the 
Wess-Zumino part. Both parts of the action are invariant under 
the rigid spacetime supersymmetry transformations, however, for 
a specific choice of normalization 
of the Wess-Zumino term, the whole action turns out to have an extra 
local symmetry known as $\kappa$-symmetry. 
The $\kappa$-symmetry allows one to remove 
the redundant fermionic degrees of freedom \cite{TOW}. 
Here we are only concerned with the bosonic part of the DBI action. 
Later on, we will explain on an alternative way to supersymmetrize the 
action using the methods of topological field theory \cite{ALI, ALI2}.  

When a constant background B-field is turned on, the only change in 
the DBI action is to replace $F$ -- the field strength of the $U(1)$ gauge 
field -- by $F+B$ everywhere in the action. Using the Seiberg-Witten map, 
one could also recast 
the action in terms of the open string variables  and  noncommutative 
filed strength ${\hat F}$ \cite{SW}. 
In this description, the effect of $B$ appears 
only through the open string metric, coupling constant, and ${\hat F}$. 
Considering a  
supersymmetric extension of the DBI action in a constant B-field, and 
looking for BPS states which preserve part of the supersymmetry leads 
to a deformed instanton equation as the BPS condition. 
These deformed equations, the so-called nonlinear instanton 
equations \cite{MOO}, are our main interest in this article. 

We start our discussion with the DBI action in the absence of any background 
field in four dimensions. By adding to the DBI action some appropriate 
terms, which are either constant or total derivatives, an action is 
derived  that has instantons as its critical points \cite{SP}. 
The structure of these extra terms is easy to guess if one notes that the 
action has to vanish at infinity where $F=0$, and the fact that it is 
defined up to some topological terms. However, when there is a background 
B-field, these terms are very nontrivial to guess. 
In this case we derive the 
action by requiring, as in the case of $B=0$, that it vanish at infinity, 
and its critical points coincide with the nonlinear instantons 
\cite{SW, MOO, T}. 
Actually the action will turn out to be proportional to the square of 
nonlinear instanton equations. This will allow us to supersymmetrize 
the action using the topological filed 
theory methods. As for the D6-branes in ${\bf R}^6$, we look for an action 
which vanishes at infinity, and has some possible extra topological terms. 
The proposed action then will have nonlinear instantons as its critical 
points.

\section{Topological DBI action in a background B-field}

In this section, we consider first the case of $B=0$ which has also 
been discussed in \cite{SP}. Next, D-branes in a constant 
background B-field are examined. In both cases, we propose actions which 
are the same as the usual DBI actions up to some constant or total 
derivative terms. The point for choosing such actions is that they can 
be written as the square of some sections which have zeros on BPS 
configurations, and can easily be supersymmetrized. 
We work out the Lagrangian in the case of $B\neq 0$, and derive 
a supersymmetric extension which has nonlinear instantons  \cite{SW} 
as the fixed points of the corresponding fermionic symmetry. 
Before doing 
any calculations, let us first  
express the result for the bosonic part of the Lagrangian explicitly  
\be
\cL = \sqrt{\det (g+M)} -\sqrt{\det(g+B)} +\f{(1-\pf B)}{\sqrt{\det(g+B)}}
(\pf F +\f{1}{4}\ep^{ijkl}B_{ij}F_{kl}) - \f{B_+^{ij}F_{ij}}{\sqrt{\det(g+B)}}
\, .
\ee
As can be seen, apart from the second term which is a constant, all extra 
terms are total derivative. These extra terms are needed if we demand that the 
Lagrangian vanish on BPS configurations. And we demand this to be able to 
supersymmetrize the action in a topological way.

To begin with,  let us get on to the case of $B=0$. Expanding 
the determinant appearing in the DBI action, we will have  
\bea
\det (g+F)& =& 1 + \det F +\f{1}{2}F^2 + 2\pf F -2\pf F \nn \\
&=& (1- \pf F)^2 + F_+^2\, , \label{re}
\eea
where $F^2=F_{ij}F^{ij}$, $F^+_{ij}=\f{1}{2}(F_{ij} 
+\f{1}{2}\ep_{ijkl}F^{kl})$,
$\pf F =\f{1}{8}\ep_{ijkl}F^{ij}F^{kl}$, and we use the flat metric 
$g_{ij}=\del_{ij}$ everywhere. Rewritting (\ref{re}), we have
\be
\left(\sqrt{\det (g+F)} +1 -\pf F\right)
\left(\sqrt{\det (g+F)} -1 +\pf F
\right)=F^2_+\, .
\label{plus}
\ee
However, in the following, we will show that the term 
\[
f=\sqrt{\det (g+F)} +1 -\pf F
\]
is positive definite. On the other hand, a natural choice for the DBI 
Lagrangian which 
vanishes at infinity (where $F=0$), and is different from the usual DBI 
Lagrangian by a constant and a total derivative is
\be
\cL = \sqrt{\det (g+F)} -1 +\pf F\, .
\ee
By looking at (\ref{plus}), we realize that choosing $\cL$ as above 
has also the advantage of being proportional 
to the square of a section (here $F^+$), and therefore can be supersymmetrized
using topological field theory methods. This will be explained shortly. 
Moreover, since $f$ is positive 
definite, then eq. (\ref{plus}) implies that 
$\cL =0$ if and only if $F_+=0$. 

To see that $f$ is positive definite, first note that it can be written as
\be
f= \sqrt{\det (g+F)} +1 -\pf F = \f{1}{2\sqrt{ \det (g+F)}}
\left\{\left(\sqrt{\det (g+F)} +1 -\pf F \right)^2 + F_+^2\right\}\, .
\label{plus2}
\ee
This equation shows that the left hand side is not negative. However, if it 
is zero, then $F_+$ must vanish, and this implies 
\[
\sqrt{\det (g+F)} +1 -\pf F = \sqrt{\det (g+F)} +1 +\f{1}{4}F^2
\]
which is always positive and cannot be zero. Thus eq. (\ref{plus2}) 
tells us  that $f$ (now call it $1/h^2$) is positive definite. 
As such, using (\ref{plus}), we can write
\[
\cL = \sqrt{\det (g+F)} -1 +\pf F = h^2 F_+^2\, .
\]   
This also proves that $\cL =0$ if and only if $F_+=0$.

To extend the above results to the case of $B\neq 0$, 
we consider the BPS condition for Euclidean 
D4-branes in a background constant B-field. Noticing 
that $F$ has to vanish at infinity, this condition reads \cite{SW}
\[
M_{ij}^+ = \f{1 -\pf M +\sqrt{\det (g+M)}}{1 -\pf B+\sqrt{\det (g+B)}}
B_{ij}^+ \equiv \al B_{ij}^+\, ,
\]
where $M= F+B$.  As in the above case with $B=0$, in the following,
 we  construct an action which has the same 
critical points as the above BPS configurations. The result will be
\bea
\cL &=& \sqrt{\det (g+M)} -\sqrt{\det(g+B)} +\f{(1-\pf B)}{\sqrt{\det(g+B)}}
(\pf F +\f{1}{4}\ep^{ijkl}B_{ij}F_{kl}) - \f{B_+^{ij}F_{ij}}{\sqrt{\det(g+B)}}
 \nn \\
 &=& \f{(M_+ -\al B_+)^2}{2\al\sqrt{\det(g+B)}}\equiv N_+^2 \, .\label{exp}
\eea
Notice that $\al$ is positive definite (for the same 
reason that the left hand side of (\ref{plus2}) is positive definite).  
Therefore $\cL$ vanishes if and only if 
$M^+=\al B^+$, i.e., it localizes on the BPS configurations.

To prove  (\ref{exp}), first note that
\be
\f{M_+^2}{B_+^2}=\left( \f{1 -\pf M +\sqrt{\det (g+M)}}{1 -
\pf B+\sqrt{\det (g+B)}}\right)\left(\f{-1 +\pf M +\sqrt{\det (g+M)}}
{-1 +\pf B+\sqrt{\det (g+B)}}\right)\equiv \al\bet \, ,
\ee
using eq.(\ref{plus}). Therefore, we can write
\bea
(M_+ -\al B_+)^2 &=& M_+^2 +\al^2 B_+^2-2\al B_{+ij}M_+^{ij} \nn \\
&= & \al\bet B_+^2 +\al^2 B_+^2 -2\al B_+^2-2\al B_{+ij}F_+^{ij} \nn \\
&= & \al B_+^2(\al + \bet  -2) -2\al B_{+ij}F_+^{ij}\, . \label{plug}
\eea
After a little algebra, the first term in the last equality can be written as
\bea
&& \al B_+^2(\al + \bet  -2)= \nn \\
&& 2\al\sqrt{\det (g+B)}\left( 
\sqrt{\det (g+M)} -\sqrt{\det(g+B)} +\f{(1-\pf B)}{\sqrt{\det(g+B)}}
(\pf F +\f{1}{4}\ep^{ijkl}B_{ij}F_{kl})\right)\, .\nn
\eea
Plugging this into eq. (\ref{plug}), finally we arrive at (\ref{exp}).

Now that $\cL$ has been written as a square of $N^+$, we employ 
the topological field theory methods  
to supersymmetrize it \cite{W}. To do so, we first introduce 
a ghost one-form field $\ps^i$, the fermionic partner of $A^i$,
 a scalar field $\ph$, and a BRST-like operator $\del$ with an action
\[
\del A^i =i\ep \ps^i \, ,\ \ \ \del\ps^i =-\ep \partial^i\ph\, \ \ \ 
\del\ph =0\, ,
\]
where $\ep$ is a constant anticommuting parameter. Further we need to 
introduce the anti-ghost fields; a self-dual 2-form $\ch_{ij}$ 
(the conjugate field to $N^+_{ij}$), a scalar $\et$, 
as well as a scalar $\la$ with ghost number 2. These tarnsform under 
$\del$ as follows
\bea
\del \ch_{ij}= \ep H_{ij}\, ,\ \ \ \del H_{ij}=0 \nn \\
\del \la =2i\ep\et\, , \ \ \ \del\et =0\, ,\nn
\eea
here the auxiliary self-dual field $H_{ij}$ has been introduced to close 
the algebra off shell.
Let us define the operator $Q$ by $\del \Phi =-i\ep\{ Q,\Phi\}$, for 
any field $\Phi$. We would like the Lagrangian to be a BRST commutator, 
i.e. $\cL_{\rm S} =i\{Q, V\}$, for some gauge fermion $V$. Since $Q^2$ acting 
on any field is zero up to a gauge transformation, this ensures that 
$\cL$ is invariant under the fermionic symmetry $Q$ if $V$ is chosen to be 
gauge invariant. A minimal choice for $V$ is 
\[
V = \ch^{ij}(H_{ij}-2N_{ij}^+) +\f{1}{2}\ps^i\partial_i\la
\]
We now vary $V$ to get $\cL$
\[
\cL_{\rm S} = i\{Q,V\}= -H^{ij}H_{ij} +2 H^{ij}N_{ij}^+ -2i\ch^{ij}
\f{\del N^+_{ij}}{\del A^k}\ps^k  
+\f{1}{2}\pl_i\ph\pl^i\la +i\ps_i\pl^i\la \, .
\]
The auxiliary field $H_{ij}$ can be integrated out using its equation of 
motion. Doing this, we obtain
\bea
\cL_{\rm S} &=& N_+^2 -2i\ch^{ij}
\f{\del N^+_{ij}}{\del A^k}\ps^k  
+\f{1}{2}\pl_i\ph\pl^i\la +i\ps_i\pl^i\la \nn \\
 &=& \sqrt{\det (g+M)} -\sqrt{\det(g+B)} +\f{(1-\pf B)}
{\sqrt{\det(g+B)}}
(\pf F +\f{1}{4}\ep^{ijkl}B_{ij}F_{kl}) - 
\f{B_+^{ij}F_{ij}}{\sqrt{\det(g+B)}}
 \nn \\ 
&-&2i\ch^{ij}
\f{\del N^+_{ij}}{\del A^k}\ps^k  
+\f{1}{2}\pl_i\ph\pl^i\la +i\ps_i\pl^i\la
\, .
\eea

\section{D6-branes in ${\bf R}^6$}

The last issue we would like to discuss is the DBI action of  flat 
Euclidean D6-branes in ${\bf R}^6$ with $B=0$. Here one may expect that,   
as in the case of topological action of D4-branes which has Yang-Mills 
instantons as its critical points, the topological 
action of D6-branes 
will have the K\"ahler-Yang-Mills instantons as its critical 
points.\footnote{ The K\"ahler-Yang-Mills equations  
\[
k_{ij}F^{ij}=0\, ,\ \ \ \ F^{2,0}=0\, ,
\]
naturally arise in the construction of a cohomological field theory on 
Calabi-Yau 3-folds \cite{JAJ}. There they appear as the fixed points of the 
corresponding BRST symmetry.} But, as 
we will see, this is not the case. Instead, the proposed topological 
Lagrangian 
\be
\cL= \sqrt{\det (g+F)} -1 + \f{1}{16}\ep^{ijklmn} k_{mn}F_{ij}F_{kl}\, ,
\label{6}
\ee 
localizes on the solutions of the following equations; 
\be
\pf F =\f{1}{2}k_{ij}F^{ij}\ \ ,{\rm and}\ \  F^{2,0}=0\, ,
\ee
where $i,j,\ldots =1,\ldots ,6$, and $k=\f{1}{2}k_{ij}dx^i\wedge dx^j$ is 
the K\"ahler form. These equations are also derived in \cite{MOO} as the BPS 
conditions for D6-branes. 
In the following, we motivate the above definition of the topological action 
for D6-branes in ${\bf R}^6$. 

To start, we use the following identity on flat Euclidean ${\bf R}^6$ 
\cite{TOW, SCH}
\be
\det (g+F) = \rho_6^2 \, ,\label{rho}
\ee
with
\[
\rho_6 = \ep^{ijklmn}\left\{ \f{1}{6!}\ep_{ijklmn}\tau_1 \ga_7 -\f{1}{2\cdot 4!}\tau_2\ga_{ijkl}F_{mn} 
+\f{i}{16}\tau_1 \ga_{ij}F_{kl}F_{mn} -\f{1}{48}\tau_2 F_{ij}F_{kl}
F_{mn}\right\}\,
,
\]
where $\tau_1, \tau_2$, and $\tau_3$ are the Pauli matrices.
Alternatively, since $\f{1}{4!}\ep^{ijklmn}\ga_{klmn}= i\ga_7\ga^{ij}$, we 
can write (\ref{rho}) as
\be
\det (g+F) = \left( \ga_7 +\f{i}{16}\ep^{ijklmn}\ga_{mn}F_{ij}F_{kl}\right)^2 
+\left(\f{i}{2}F^{ij}\ga_7\ga_{ij} +\pf F \right)^2 \, .\label{th}
\ee

Now take $\th$ to be a constant commuting left-handed spinor on ${\bf R}^6$. 
We choose the gamma matrices to be hermitian and antisymmetric, and 
in the complex coordinate let $\ga_\al\th =0$, with the normalization 
$\th^\dagger\th =1$. The K\"ahler form and the 
holomorphic 3-form are then defined as follows \cite{JAJ}
\bea
&& k_{ij}= i\th^{\dagger}\ga_{ij}\th \nn \\
&& C_{ijk}= \th^\dagger \ga_{ijk}\th^*\, . \label{K}
\eea

Expanding eq. (\ref{th}), and multiplying it by $\th^\dagger$ on left and 
$\th$ on right,  
and using (\ref{K}) results in the following identity 
\be
\det (g+F) = \left( 1 -\f{1}{16}\ep^{ijklmn} k_{mn}F_{ij}F_{kl}\right)^2 
+\left(\pf F -\f{1}{2}F^{ij} k_{ij} \right)^2 
+2F_{\al\bet}F^{\al\bet} +\f{1}{2}\tf_{\al\bet}\tf^{\al\bet} \, ,\label{alpha}
\ee
where $\al ,\bet ,\ga \ldots $ are the complex holomorphic tangent indices, 
$\tf^{ij}=\f{1}{4}\ep^{ijklmn}F_{kl}F_{mn}$, and use has been made of
\be
\f{1}{4}(k_{ij}F^{ij})^2 =\f{1}{8}\ep^{ijklmn}F_{ij}F_{kl}k_{mn} +
\f{1}{2}F^2 -2F_{\al\bet}F^{\al\bet}\, ,\label{id}
\ee
for any rank 2 antisymmetric tensor in 6 dimensions. 
Let us now write (\ref{alpha}) as
\bea
&& \left(\sqrt{\det (g+F)} -1 + \f{1}{16}\ep^{ijklmn} k_{mn}F_{ij}F_{kl}
\right)
\left(\sqrt{\det (g+F)} +1 - \f{1}{16}\ep^{ijklmn} k_{mn}F_{ij}F_{kl}
\right)\nn \\
 &=& \left(\pf F -\f{1}{2}F^{ij} k_{ij} \right)^2 
+2F_{\al\bet}F^{\al\bet} +\f{1}{2}\tf_{\al\bet}\tf^{\al\bet} \, .\label{al}
\eea 

As in the case of D4-branes, first we prove that the term 
\[
{\tilde f}=\sqrt{\det (g+F)} +1 -\f{1}{16}\ep^{ijklmn} k_{mn}F_{ij}F_{kl}
\]
on the left hand side of eq. (\ref{al}) is positive definite. 
To show this, notice that the left hand 
side of eq. (\ref{al}) can be written as 
\bea
&& \left(\sqrt{\det (g+F)} +1 - \f{1}{4}k^{ij}\tf_{ij}
\right)
\left(\sqrt{\det (g+F)} -1 + \f{1}{4}k^{ij}\tf_{ij}\right)
\nn \\
&=& -\left(\sqrt{\det (g+F)} +1 - \f{1}{4}k^{ij}\tf_{ij}
\right)
\left(\sqrt{\det (g+F)} +1 - \f{1}{4}k^{ij}\tf_{ij}
-2\sqrt{\det (g+F)}\right)\nn \\
&=& -\left(\sqrt{\det (g+F)} +1 - \f{1}{4}k^{ij}\tf_{ij}
\right)^2
+2\sqrt{\det (g+F)} \left(\sqrt{\det (g+F)} +1 - \f{1}{4}k^{ij}
\tf_{ij}\right)\nn\, .
\eea
So, using (\ref{al}), we can write
\bea
&& 2\sqrt{\det (g+F)} \left(\sqrt{\det (g+F)} +1 - \f{1}{16}\ep^{ijklmn}
k_{mn}F_{ij}F_{kl}\right) \nn \\
&=& \left(\sqrt{\det (g+F)} +1 - \f{1}{16}\ep^{ijklmn}
k_{mn}F_{ij}F_{kl}\right)^2 + \left(\pf F -\f{1}{2}F^{ij} k_{ij} \right)^2\nn \\ 
&+& 2F_{\al\bet}F^{\al\bet} +\f{1}{2}\tf_{\al\bet}\tf^{\al\bet} \, .
\label{so}
\eea 
Therefore ${\tilde f}$ is not negative. But if it is zero, according to the above 
equation we must have
\[
\pf F=\f{1}{2}k_{ij}F^{ij}\, ,\ \ \ F_{\al\bet}=0\, ,
\]
(equation $\tf_{\al\bet}=0$ is automatically satisfied when
$F_{\al\bet}=0$). 
So eq. (\ref{alpha}) becomes
\be
\det (g+F) = \left( 1 -\f{1}{16}\ep^{ijklmn} k_{mn}F_{ij}F_{kl}\right)^2 \, ,
\ee
we write this as
\bea
\f{1}{8}\ep^{ijklmn} k_{mn}F_{ij}F_{kl} &=& 
-\det (g+F) + 1 +\left(\f{1}{16}\ep^{ijklmn} k_{mn}F_{ij}F_{kl}\right)^2\nn \\
&=& -1 -\det F -\f{1}{2}F^2 -\f{1}{8}\tf^{ij}\tf_{ij} 
+1 +\f{1}{8^2}\left(\f{1}{8} \ep^{ijklmn} k_{mn}F_{ij}F_{kl}+
8\tf^{ij}\tf_{ij}\right)
\nn \, , \label{af}
\eea
where in the last line we have expanded the determinant, and used 
(\ref{id}) with $\tf_{\al\bet}=0$. Finally the above equation implies that 
\[
(1 -\f{1}{8^2})\f{1}{8}\ep^{ijklmn} k_{mn}F_{ij}F_{kl}= -\det F -\f{1}{2}
F^2 < 0 \, .
\]
Hence 
\[ 
\tilde{ h}^{-2}\equiv \tilde {f}=\sqrt{\det (g+F)} +1 -\f{1}{16}\ep^{ijklmn} 
k_{mn}F_{ij}F_{kl}
\]  
cannot vanish and is positive definite.

The above calculations show that a good choice for the DBI Lagrangian of 
D6-branes in ${\bf R}^6$ is the one in (\ref{6}). 
This is justified for the following reasons. Firstly, it vanishes 
at infinity where $F=0$, and up to a constant and a total derivative 
is the usual DBI Lagrangian. Secondly, eq. (\ref{al}) 
and the positive definiteness of ${\tilde f}$ allow us to write (\ref{6}) as 
\bea
\cL &=& \sqrt{\det (g+F)} -1 + \f{1}{16}\ep^{ijklmn} k_{mn}F_{ij}F_{kl}\nn \\
&= &{\tilde h}^2\left\{ \left(\pf F -\f{1}{2}F^{ij} k_{ij} \right)^2 
+2F_{\al\bet}F^{\al\bet} +\f{1}{2}\tf_{\al\bet}\tf^{\al\bet}\right\} \, ,
\label{R}
\eea
which shows that $\cL$ is zero 
if and only if $\pf F=\f{1}{2}k_{ij}F^{ij}\, ,\ F_{\al\bet}=0$. 

Since the right hand side of (\ref{R}) is the sum of the squares of 
the sections 
\bea
&& s^{(1)}= {\tilde h}\left(\pf F -\f{1}{2}F^{ij} k_{ij} \right)\nn \\
&& s^{(2)}_{\al\bet}={\tilde h}F_{\al\bet}\nn \\
&& s^{(3)}_{\al\bet}={\tilde h}\tf_{\al\bet}\nn\, ,
\eea
we can employ the same method that we used in the previous section to 
supersymmetrize the action. For the case of K\"ahler-Yang-Mills 
equations, this has been explicitly done in \cite{JAJ}.

\pagebreak

\end{document}